\documentclass[conference]{IEEEtran}
\IEEEoverridecommandlockouts
\usepackage{cite}
\usepackage{amsmath,amssymb,amsfonts}
\usepackage{algorithmic}
\usepackage{graphicx}
\usepackage{textcomp}
\usepackage{xcolor}
\usepackage{caption}
\def\BibTeX{{\rm B\kern-.05em{\sc i\kern-.025em b}\kern-.08em
    T\kern-.1667em\lower.7ex\hbox{E}\kern-.125emX}}
\begin{document}

\title{Advancements in Non-Invasive Neuroimaging: Exploring the Potential of Radar Technology for Brain Imaging and Tumour Detection}


\author{
    \IEEEauthorblockN{Keniel Peart, Indu Bodala, Shelly Vishwakarma}
    \IEEEauthorblockA{Department of Electronic and Computer Science, University of Southampton, UK}
         [k.peart, i.p.bodala, s.vishwakarma]@soton.ac.uk
}

\maketitle

\thanks{\textcopyright~2024 IEEE. Personal use of this material is permitted. Permission from IEEE must be obtained for all other uses, in any current or future media, including reprinting/republishing this material for advertising or promotional purposes, creating new collective works, for resale or redistribution to servers or lists, or reuse of any copyrighted component of this work in other works. Published in 2024 46th Annual International Conference of the IEEE Engineering in Medicine and Biology Society (EMBC). DOI: 10.1109/EMBC53108.2024.10782724}

\begin{abstract}
This study investigates radar technology for non-invasive brain imaging and tumour detection, offering an alternative to MRI and CT scans. Using Ansys HFSS to simulate electromagnetic interactions in brain tissues, we evaluate the penetration, signal strength, and safety of Patch and Vivaldi antennas. Results show Patch antennas are optimal for tumour localization, while Vivaldi antennas suit broader scanning applications. Although promising for safer, more accessible imaging, especially in resource-limited environments, further research with diverse models and actual patient data is essential to advance this technology in non-invasive medical diagnostics.
\end{abstract}


\section{Introduction}
\label{sec:intro}

Advancements in brain imaging techniques such as Magnetic Resonance Imaging (MRI), Functional MRI (fMRI), Computed Tomography (CT), Positron Emission Tomography (PET), and Electroencephalography (EEG) have significantly contributed to our understanding of the complex operations of the human brain, understanding its functions, and various disorders.\cite{westerlund_stem_2003}. However, despite their advancements, these methods face limitations including high cost, radiation risks, and accessibility issues, underscoring the need for more innovative, accessible, and non-invasive brain imaging solutions, particularly in diverse and resource-limited settings.

Radar sensing already successful in multitude of domains can be a safer and more accessible alternative to traditional radiation-based brain imaging techniques like CT and PET scans. Utilizing Ultra-Wideband (UWB) antennas, radar sensing can offer effective penetration and clearer imaging of the brain, especially in tumour detection \cite{hossain_lightweight_2023}. Innovations in antenna design, such as elliptical and brick shapes, have enhanced signal transmission and reception through the head, contributing significantly to the field \cite{talukder_compact_2021, rodriguez-duarte_brick-shaped_2020}. This technology is also showing promise in detecting strokes and brain injuries, analyzing the response of different brain tissues to radar signals \cite{ullah_experimental_2022}.  Deep-learning can further accelerate this growth. Models like MBINet and BrainImageNet have already been demonstrated to be instrumental in the classification and segmentation of brain tumours and can be potentially used in radar based imaging methods\cite{hossain_brain_2023}. Some researches have used techniques like pattern-re-configurable metasurface antenna arrays to observe brain functions more accurately  \cite{ ojaroudi_pattern-reconfigurable_2021,akazzim_multi-element_2022, ghavami_use_2022}. Although significant progress has been achieved, much of this research continues to depend on the use of realistic head models and simulation software like HFSS to guarantee safety and accuracy in imaging through simulated testing \cite{ margish_s_joshi_analysis_2016}.

Building upon this foundational work, our research aims to assess the viability of radar technology for brain imaging by addressing two pivotal questions: first, can electromagnetic waves effectively penetrate head tissue, and second, can radars reliably detect tumors within the brain? Furthermore, this paper presents a unique investigation into the feasibility and safety of radar technology for brain imaging, with a particular emphasis on comparing the electromagnetic penetration capabilities of Patch and Vivaldi antennas. What sets this research apart is its simultaneous examination of the effectiveness and stability of these antennas within a single head model study while also exploring the intricate relationship between electromagnetic penetration and tumor detection. Our study not only scrutinizes how these antennas penetrate tissue and detect brain anomalies but also explores how the anomalies themselves influence the penetration and imaging process. This approach signifies a significant advancement in radar brain imaging research, bridging existing gaps in the field and offering valuable insights into optimizing the safety and diagnostic accuracy of this emerging technology.

The paper is organized as follows: Section \ref{sec:expMethods} provides an in-depth look at our methodology, including details on the simulation, antenna design, and experimental setup. Sections \ref{sec:experiment1} and \ref{sec:experiment2} discusses the results and findings from our experiments. Finally, Section \ref{sec:conclusions} concludes the paper, reflecting on the insights obtained and future prospects in the application of radar technology in medical diagnostics and brain research.

\section{Simulation Methodology}
\label{sec:expMethods} 
In this section, we provide a comprehensive overview of our simulation methodology, with a primary focus on investigating how electromagnetic waves interact with head tissues, including their behavior and responses to irregularities such as tumours. Our approach employs Ansys HFSS to build a highly detailed head model and develop the electromagnetic signal framework. To create an effective electromagnetic signal model, we conduct simulations involving two types of antennas: the Patch Antenna and the Vivaldi Antenna. We assess the performance of these antennas using a predefined set of metrics, which will be discussed in subsequent sections.

\subsection{Human Head Modelling}
\begin{figure}[h]
    \centering
    \includegraphics[width=.4\textwidth]{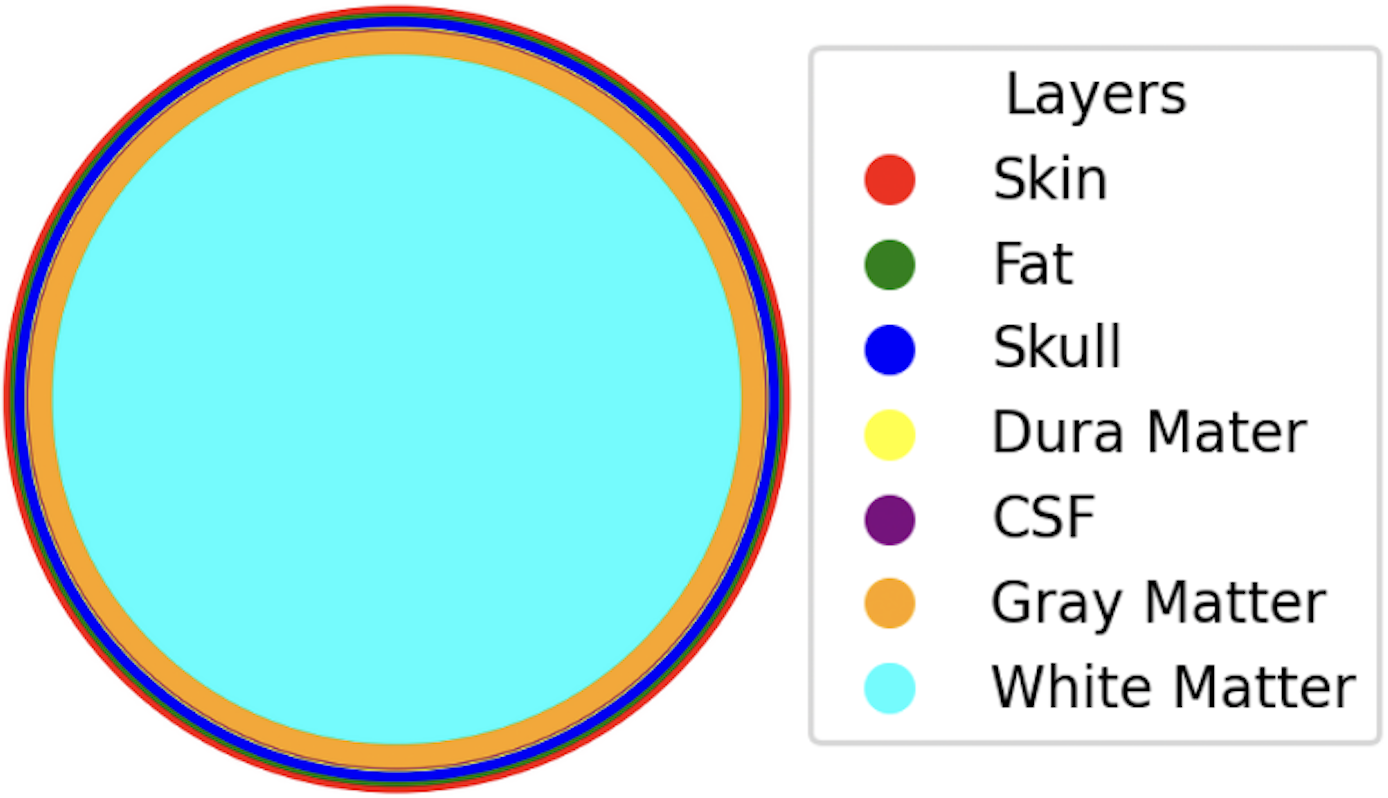} 
    \caption{Head Model Schematic}
    \label{fig:head_model_schematic} 
   
\end{figure}

We utilize Ansys HFSS, with its 3D simulation capabilities and precise meshing techniques, to create a human head model. This model consists of seven concentric spheres, each representing a distinct cranial tissue layer, enabling us to accurately emulate their dielectric characteristics. These layers comprise Skin, Fat, Skull, Dura Mater, Cerebrospinal Fluid (CSF), Gray Matter, and White Matter, as illustrated in Fig \ref{fig:head_model_schematic}. To determine the dielectric properties of the head model, we draw inspiration from Gabriel et al.'s research \cite{gabriel_dielectric_1996-1}. Additionally, we incorporate the Cole-Cole model \cite{hesabgar_dielectric_2017} to precisely characterize the frequency-dependent dielectric properties of biological tissues at 1 GHz. Detailed conductivity and relative permittivity values are provided in Table \ref{table:anatomical_structure}.

\begin{table}[h]
\centering
\begin{tabular}{|l|c|c|c|c|}
\hline
\textbf{Layer} & \textbf{R (mm)} & \textbf{Depth (mm)}  & \textbf{$\epsilon$} & \textbf{ $\sigma$}  \\
\hline
Skin & 120  & 1.35 \cite{Chopra2015} & 40.93   & 0.89\\
\hline
Fat & 118.65  & 1.4 & 5.44  & 0.05 \\
\hline
Skull & 117.25  & 5.3 \cite{MoreiraGonzalez2006} & 12.36  & 0.15\\
\hline
Dura Mater & 111.95  & 0.36 \cite{Fam2018} & 44.201  & 0.99\\
\hline
CSF & 111.59  & 2.1 \cite{Haeussinger2011} & 68.43  & 2.45\\
\hline
Gray Matter & 109.49 & 3.37 \cite{Winkler2010,Schnack2014} & 52.28  & 0.98\\
\hline
White Matter & 106.12 & Inner part & 38.57  & 0.62 \\
\hline
\end{tabular}
\caption{Anatomical Structure and Dielectric Properties (where $\epsilon$ and $\sigma$ stand for Relative Permittivity and Conductivity respectively) of the Simulated Head Model}
\label{table:anatomical_structure}
\end{table}

\subsection{Electromagnetic Signal Model}
To simulate the propagation of electromagnetic waves through our head model, we employ Ansys HFSS, which allows us to analyze the behavior of these waves in detail. Our methodology incorporates adaptive meshing, dynamically adjusting mesh density based on field gradients to balance computational efficiency and simulation accuracy. The iterative solver is also crucial, continuously refining electromagnetic field calculations to account for complex wave interactions with various tissue structures. Within HFSS, we utilize the Quasi-Newton method to calibrate key simulation parameters, such as frequency sweep count and convergence criteria for the iterative solver. This calibration minimizes deviations between our simulations and real-world behaviors, enhancing the precision and reliability of our methodology for understanding electromagnetic wave penetration of head tissues and interaction with tumors.

In our simulations, we utilize an Ultra-wideband electromagnetic signal source with a center frequency of 2.45 GHz, covering a frequency range from 0.5 GHz to 5 GHz. This wide spectrum is essential for examining signal interactions across different tissue densities, which is crucial for tumor detection and characterization.

\subsection{Simulation Setup}

\begin{figure}[h]
    \centering
    \includegraphics[width=.2\textwidth]{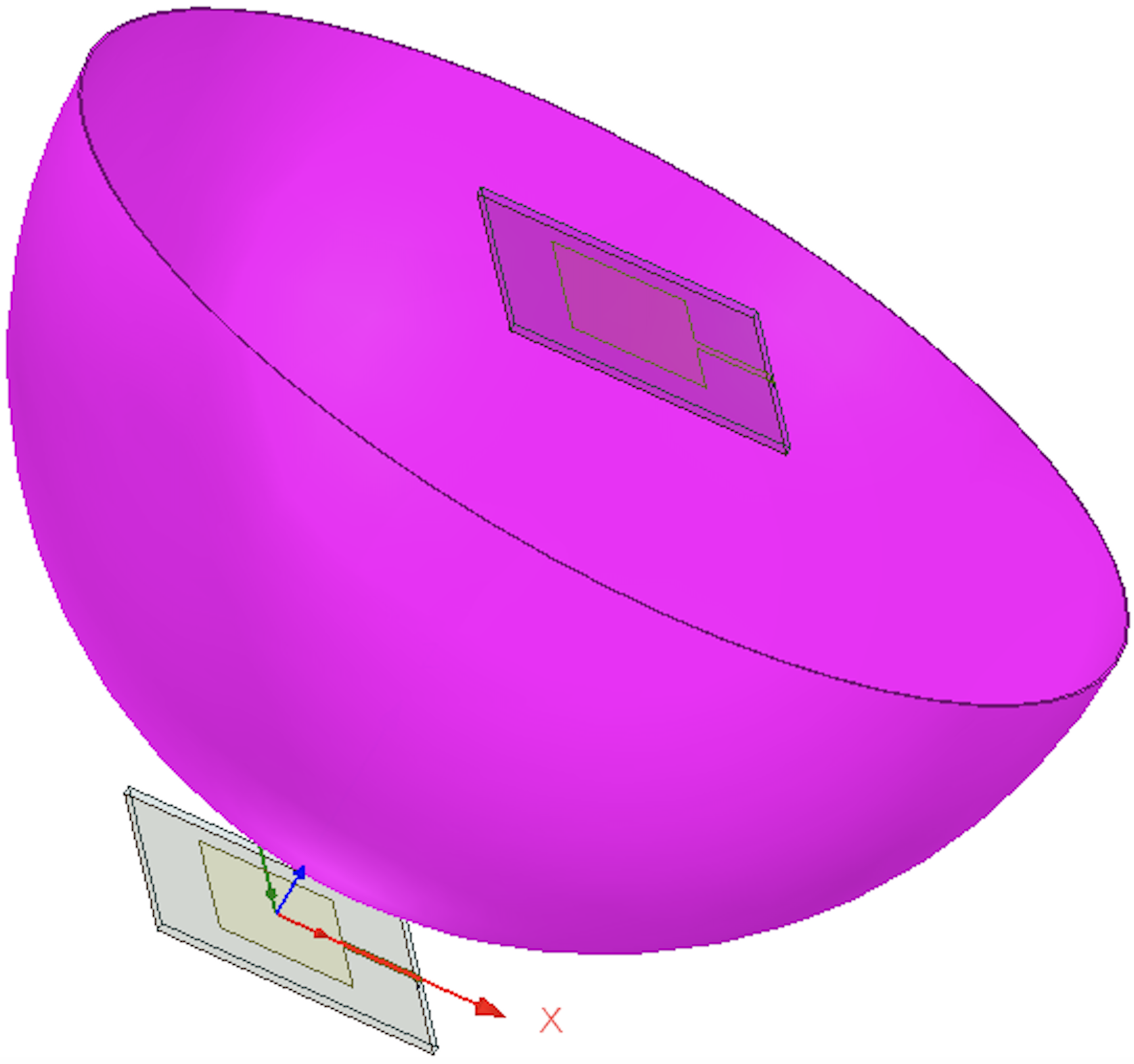} 
    \caption{Transmitting and Receiving Antennas}
    \label{fig:forward_scatter} 
   
\end{figure}

In this section, we investigate the transmission of electromagnetic waves and their interaction with brain tissues. Our simulation setup involves two antennas: one for transmitting the electromagnetic wave and another for receiving the reflected signal from the brain tissues. These antennas are arranged in the forward scatter configuration, as shown in Figure \ref{fig:forward_scatter}. We maintain a fixed distance of 10mm between the head model and the antennas.

To analyze the distribution and strength of transmitted electric fields and their interaction with the head model, we employ the Electric Field Pattern visualization technique. To assess safety, we calculate the Specific Absorption Rate (SAR) to measure tissue energy absorption. Additionally, we consider two key antenna design metrics: Return Loss, which quantifies power loss due to reflections (lower return loss indicates better antenna performance), and Antenna Efficiency (VSWR), where a lower VSWR indicates higher efficiency.

For a comprehensive study of electromagnetic wave interactions with human tissue using the ultra-wideband signal model, we have developed two specialized antennas: a Vivaldi antenna and a Patch antenna, as detailed in the following section.
\subsubsection{Antenna Module}
\begin{figure}[!htbp]
    \centering
    \begin{minipage}{0.5\textwidth}
        \centering
        \includegraphics[width=\textwidth]{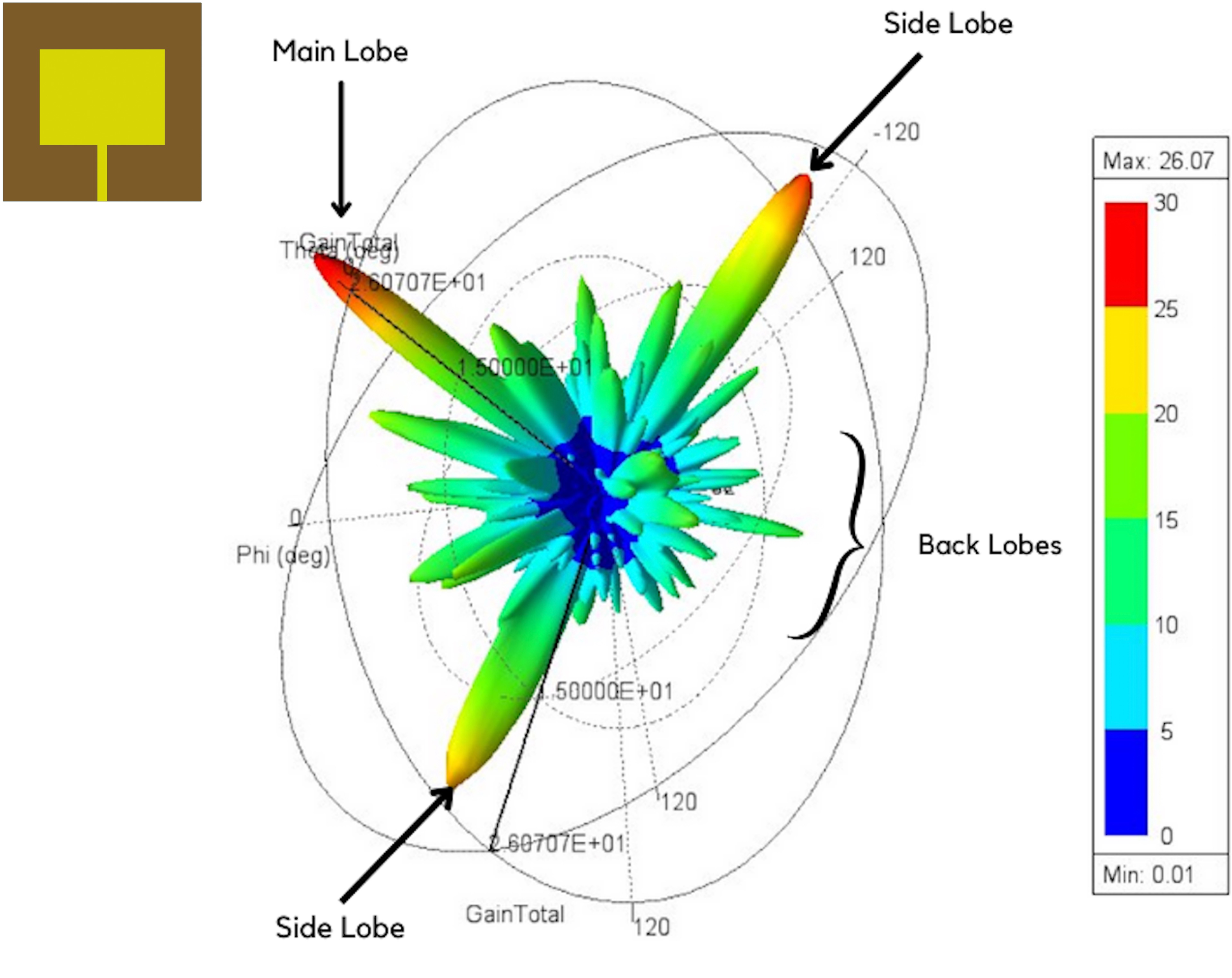} 
        \caption*{(a)}
    \end{minipage}
    \hfill
    \begin{minipage}{0.5\textwidth}
        \centering
        \includegraphics[width=\textwidth]{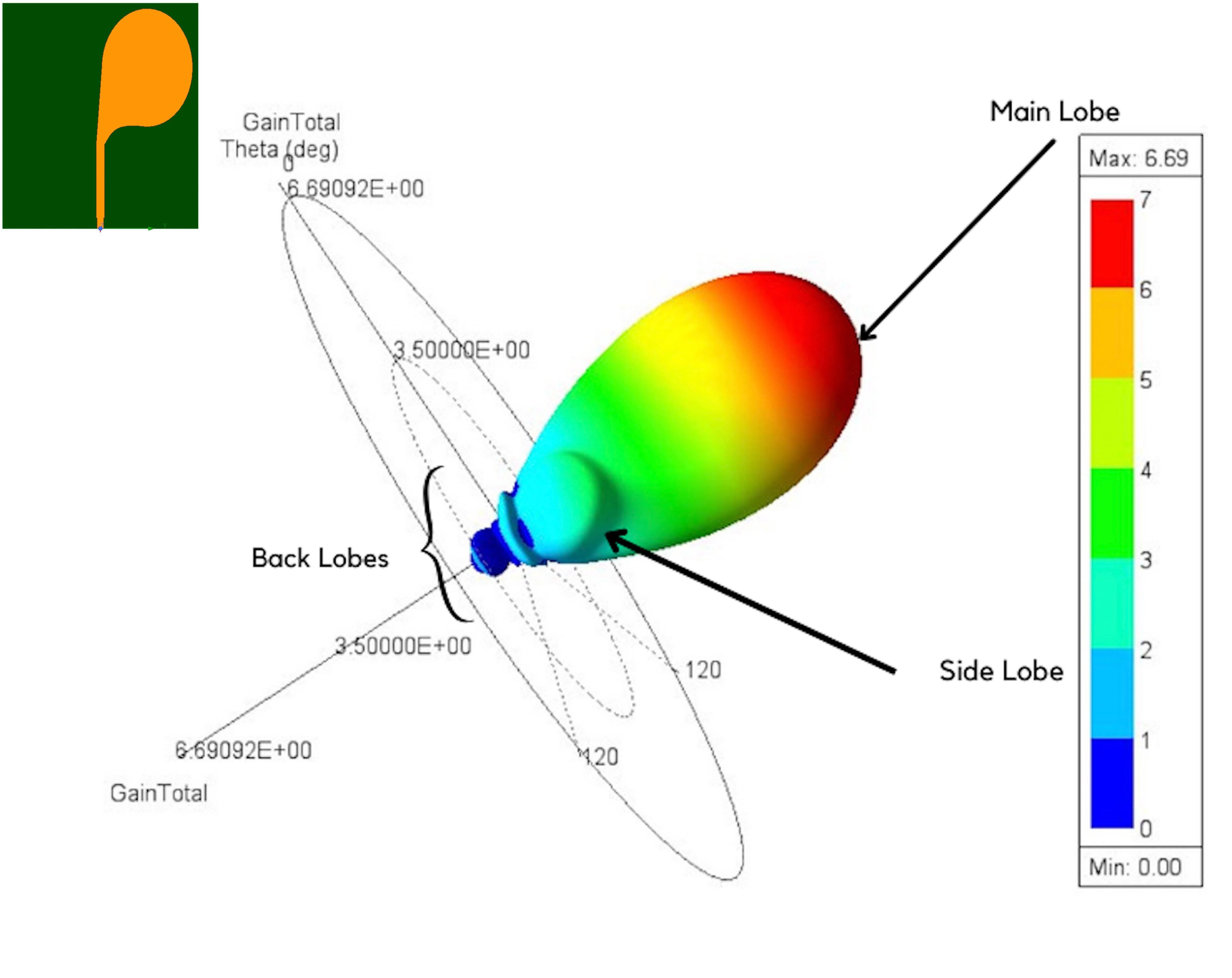} 
        \caption*{(b)}
    \end{minipage}
    \captionsetup{justification=centering}
    \caption{Schematic and Antenna Radiation Pattern: (a) patch antenna, (b)  vivaldi antenna}
    \label{fig:antenna_rad_pat}
\end{figure}

Figure \ref{fig:antenna_rad_pat} top right displays two specialized antennas, selected for their simplicity and unique properties aligned with the objectives of our study.


The Vivaldi antenna, known for its ultra-wideband capabilities and broad radiation pattern (Fig \ref{fig:antenna_rad_pat}b), plays a crucial role in accommodating experiments across a wide spectrum. It provides a comprehensive view of brain activities, meticulously calibrated to minimize interference within the 0.5 to 5 GHz frequency range. In contrast, the Patch antenna exhibits a flat, directional radiation pattern and a narrower bandwidth (Fig \ref{fig:antenna_rad_pat}a). Its rectangular design and specific frequency range result in peak efficiency at a resonant frequency, compensating for its limited bandwidth.

To evaluate the performance of these antennas, we connected the signal source (as previously described) to the antennas and allowed them to propagate electromagnetic waves into free space. This enabled us to analyze various parameters, including reflections (S11), VSWR, radiation patterns, and gain. These findings provide insights into the strengths and weaknesses of each antenna in various aspects of our study.

\section{Results and Analysis}
We assess the propagation of electromagnetic waves through the human head using both antennas in two distinct experiments to evaluate their suitability within the 0.5 to 5 GHz frequency range. These experiments focus on Electromagnetic Wave Penetration and Tumour Detection.

\subsection{Electromagnetic Wave Penetration}
\label{sec:experiment1}
\begin{figure*}[h]
    \centering
    \includegraphics[width=1\textwidth]{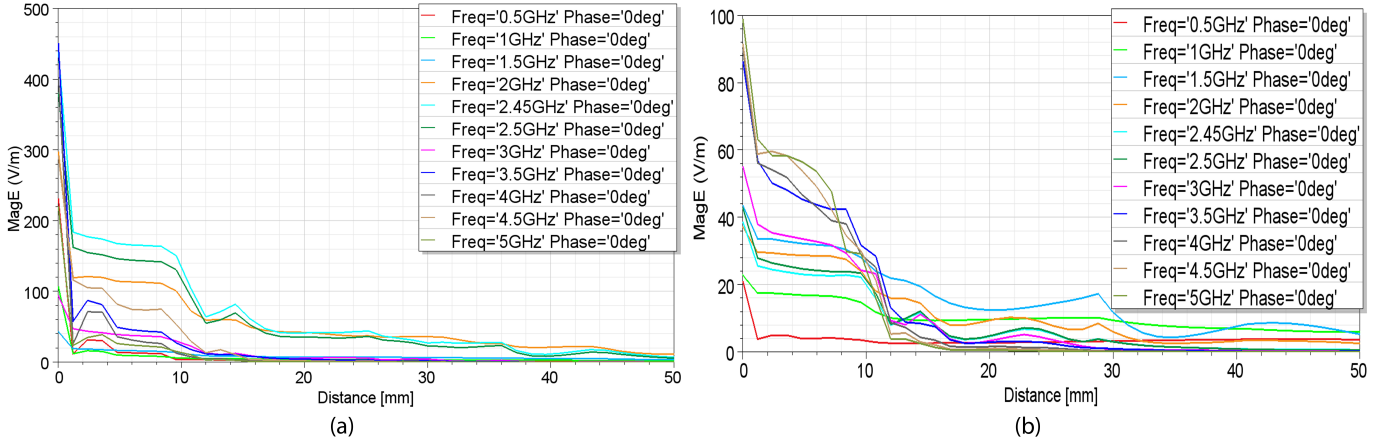} 
    \caption{Electric Field vs Distance Plot (a) Patch Antenna (b) Vivaldi Antenna)}
    \label{fig:efield_distance} 
\end{figure*}

\begin{figure*}[h]
    \centering
    \includegraphics[width=1\textwidth]{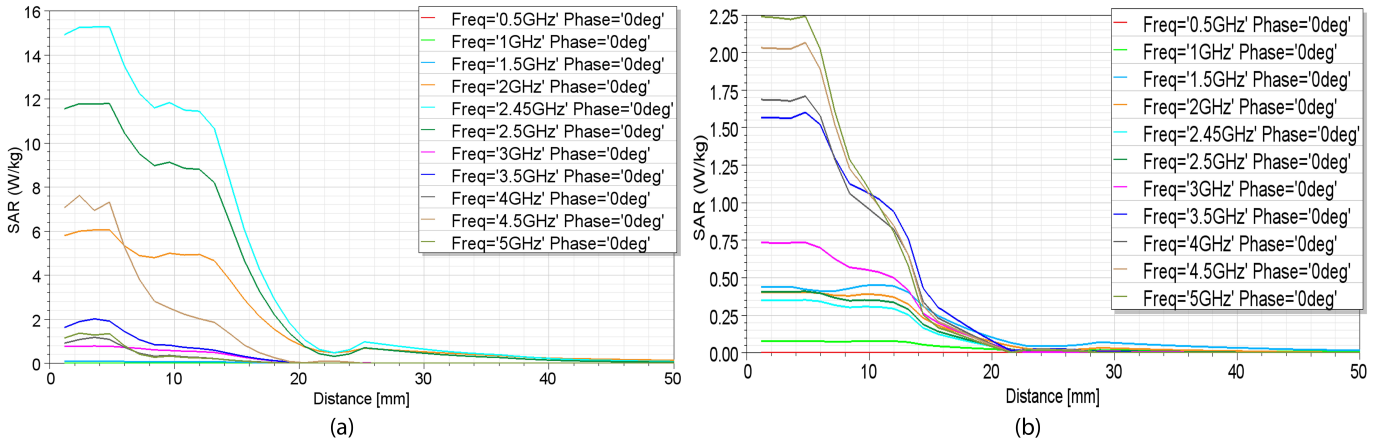} 
    \caption{SAR vs Distance Plot (a) Patch Antenna (b) Vivaldi Antenna}
    \label{fig:sars_distance} 
\end{figure*}
We begin our analysis by examining electromagnetic penetration for both antennas, as depicted in Fig \ref{fig:antenna_rad_pat}. Specifically, we focus on how the electric fields and SAR values are influenced as the waves penetrate the head model. We generate two plots: an Electric Field vs. Distance plot and a SAR vs. Distance plot. Our interest lies in understanding how these parameters change as the waves traverse the model, passing through the layers at depths indicated in Table \ref{table:anatomical_structure}. In our results, we restrict the distance of our plots to 50mm, as this allows for comprehensive penetration through all layers of the model.

\subsubsection{\textbf{Electric Field Magnitude}}
Figure \ref{fig:efield_distance} shows plots of the electric field strengths of the patch antenna compared to the vivaldi antenna as the waves penetrate the head model. 

The Patch antenna initially exhibits a robust electric field strength, as illustrated in figure \ref{fig:efield_distance}(a), measuring approximately 450 mV/m at 2.45 GHz. However, this strength rapidly diminishes within the first 10 mm, dropping below 140 mV/m. This signifies a swift decline in energy penetration through the medium. This pattern persists across various frequencies, but at higher frequencies, such as 4.5 GHz, the field strength decreases even more rapidly, reaching approximately 40 mV/m at the same distance. This suggests that the Patch antenna may not be suitable for applications requiring deep electromagnetic interaction due to its limited effective range.

In contrast, the Vivaldi antenna commences with a lower field strength, as depicted in figure \ref{fig:efield_distance}(b), measuring around 100 mV/m at 4.5 GHz. However, its strength diminishes more gradually, maintaining a strength of over 25 mV/m up to 10 mm. The field strengths of the Vivaldi antenna at different frequencies, ranging from 2.45 GHz to 4.5 GHz, exhibit greater similarity, indicating a more consistent ability to penetrate over distance. This becomes apparent when assessing the field strength at 30 mm. The Patch antenna's field strength undergoes significant fluctuations with varying frequencies, almost reaching 0 mV/m. In contrast, the Vivaldi antenna maintains a field strength of approximately 5 mV/m across most frequencies. These observations suggest that the Patch antenna is better suited for focused, high-intensity applications, while the Vivaldi antenna's consistent field strength over distance and frequency renders it more suitable for applications requiring reliable wideband performance.
\begin{figure}[htbp]
    \centering
    \includegraphics[width=.5\textwidth]{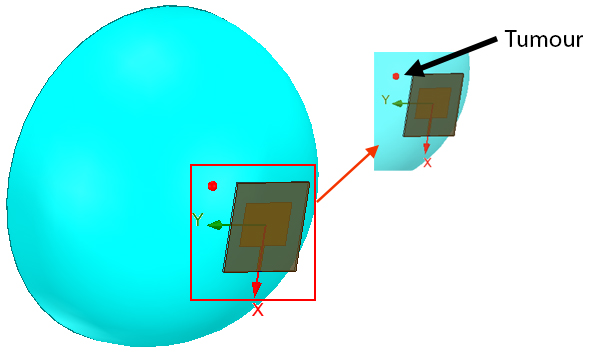} 
    \caption{Simulation setup for tumour Detection Experiment}
    \label{fig:tumoursetup} 
\end{figure}

\begin{table}[h]
\centering
\begin{tabular}{|l|c|}
\hline
\textbf{Radius (mm)} & 5 \\
\hline \textbf{Permittivity ($\epsilon$)} &  55 \\ \hline  \textbf{Conductivity ($\sigma$)} & 7 \\
\hline
\end{tabular}
\caption{Size and dielectric properties of tumour}
\label{table:tumour_dielectrics}
\end{table}

\begin{figure*}[htbp]
    \centering
    \includegraphics[scale=0.13]{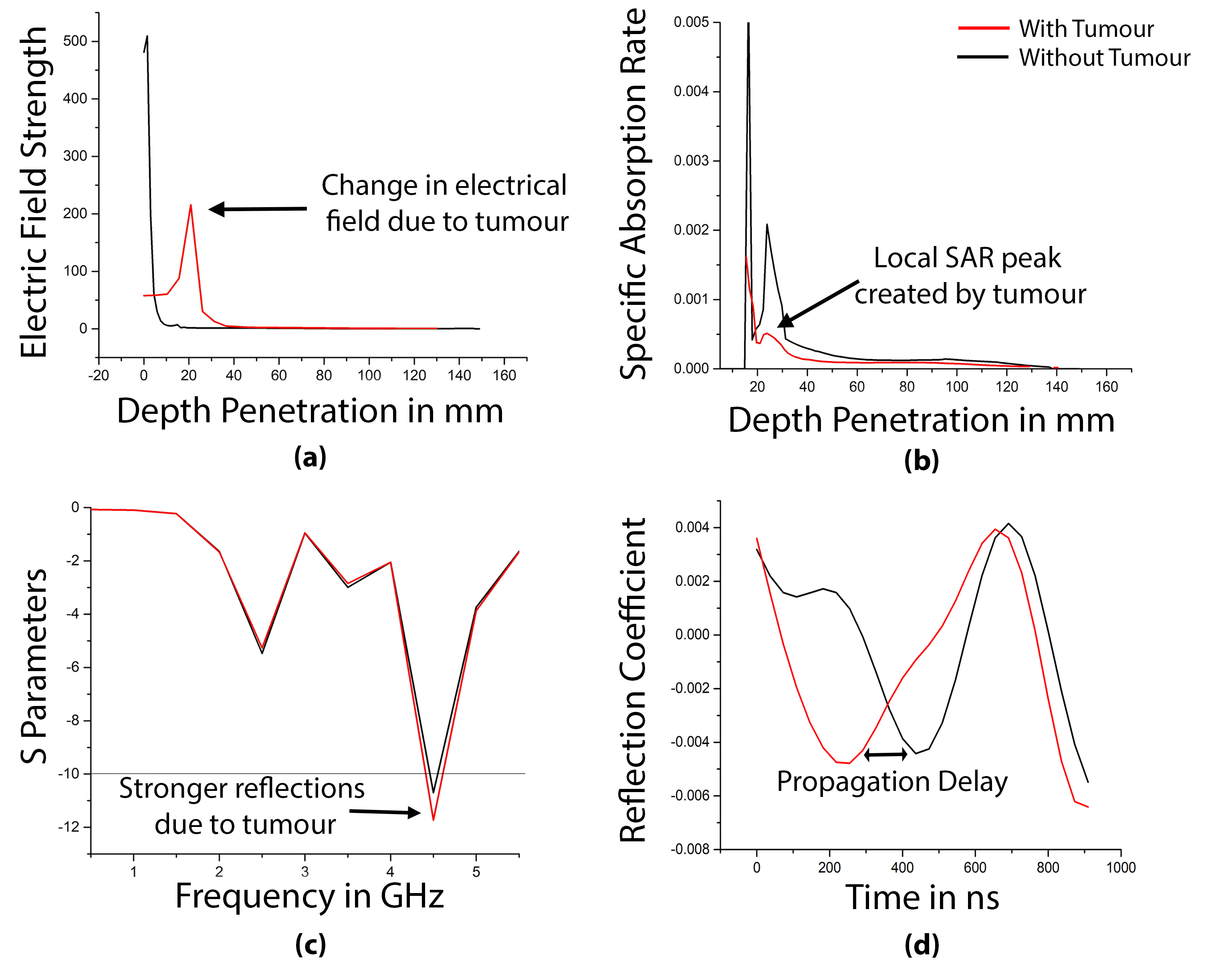} 
    \caption{ Results (a) Patch Antenna Electric Field Strength plot (b) SAR plot with and without tumour (c) Patch Antenna Reflection Coefficient plot (d) Propagation Delay with and without tumour}
    \label{fig:tumour_results} 
\end{figure*}

\subsubsection{\textbf{Specific Absorption Rates (SARs)}}
Figure \ref{fig:sars_distance} shows plots of the SARs values of the patch antenna compared to the vivaldi antenna as the waves penetrate the head model. 

The Patch antenna exhibits a significantly steeper SAR decrease (as depicted in figure \ref{fig:sars_distance}(a)) compared to the Vivaldi antenna. For instance, at 2.45 GHz, the Patch antenna starts with a high SAR of approximately 15.5 W/kg, which rapidly diminishes to around 1.5 W/kg within just 20 mm. This sharp decline indicates that energy is primarily concentrated near the antenna, potentially limiting its depth penetration while reducing the risk of affecting deeper tissues.

In contrast, the Vivaldi antenna initiates with a lower SAR (shown in figure \ref{fig:sars_distance}(b)), approximately 2.25 W/kg at the same frequency, and experiences a more gradual decrease. It remains below 0.5 W/kg even beyond 30 mm, signifying a more evenly distributed energy pattern that could facilitate deeper penetration without substantial localized heating.

At higher frequencies, such as 4.5 GHz, these distinctions become even more pronounced. The Patch antenna starts with a SAR of about 7.5 W/kg and exhibits a rapid decrease, indicating its suitability for applications requiring intense energy in close proximity to the antenna. On the other hand, the Vivaldi antenna, beginning just above 2.0 W/kg, demonstrates consistent and low absorption over a wider area, making it suitable for applications demanding uniform energy distribution with minimal impact on biological tissues.

In summary, the Patch antenna, with its focused frequency response and narrow bandwidth, proves particularly effective for precision tasks like tumor detection analysis. Its narrow-band nature may miss certain frequency-specific details, but this characteristic is advantageous for specialized applications. Conversely, the Vivaldi antenna excels in ultra-wideband (UWB) applications, such as penetration depth analysis, where a broad frequency range is essential. However, due to its complex design and susceptibility to interference, careful calibration is needed to fully utilize its extensive frequency coverage \cite{zhang_microwave_2012}.

\subsection{Tumour Detection}
\label{sec:experiment2}
Building on our previous findings that favored the Patch antenna for tumor detection, we focused our simulations exclusively on this antenna type. To simulate the presence of a tumor, we made modifications to our model by inserting a sphere that represents the tumor between the Dura and CSF layers, as illustrated in Figure \ref{fig:tumoursetup}. This sphere was designed to replicate the dielectric properties typically found in tumors, as detailed in Table \ref{table:tumour_dielectrics}. In addition, we introduced a new metric called "propagation delay" to measure the time difference between wave transmission and reception at the receiver. This metric provided valuable insights into how tumors influence the behavior of electromagnetic waves. Our methodology involved conducting two critical analyses: one with the tumor and one without. This approach allowed us to isolate and examine the specific effects of tumors on electromagnetic wave interactions.

\subsubsection{\textbf{Propagation Delay}} Our experiments clearly demonstrated that tumors had a significant impact on the propagation of electromagnetic waves, leading to changes such as delays, alterations in reflection behavior, and shifts in electric field patterns. These changes were evident in variations in the reflection coefficient (S11 parameter) and a reduction in the transmission coefficient (S21 parameter), indicating increased wave absorption and scattering due to the presence of the tumor. We also observed fluctuations in electric field strength, an approximate 200-nanosecond delay, and a phase shift in reflected signals, all of which were attributed to the influence of the tumor (as depicted in Fig. \ref{fig:tumour_results} (d)). Furthermore, changes in the Specific Absorption Rate (SAR) within the head model emphasized how tumors modified energy absorption rates at various tissue depths.
   

\subsubsection{\textbf{Reflection Coefficient}} The observed changes in electromagnetic wave behavior, particularly in terms of reflections (illustrated in Fig. \ref{fig:tumour_results} (c)), offer insights into how tumours can influence RF power reflection. In specific frequency ranges, we observe that the red line (With tumor) exhibits more pronounced dips compared to the black line (Without tumor), indicating stronger reflections or a less favorable impedance match in those instances. This phenomenon likely occurs because the electrical properties of the tumor modify the impedance of surrounding tissues, thereby affecting the overall RF system matching.

   

\subsubsection{\textbf{Electric Field Strength}}
The electric field strength (as shown in Fig. \ref{fig:tumour_results} (a)) exhibits a clear peak in the "With tumor" curve. However, this peak is delayed in comparison to the "Without tumor" curve, suggesting that the presence of a tumor impacts the penetration of the electric field. This delay can be attributed to the distinct electrical properties of the tumor, including permittivity and conductivity, which differ from those of the surrounding tissue. These differing properties influence the propagation of the electric field through the tumor.

  

\subsubsection{\textbf{Specific Absorption Rates (SARs)}} The observed Specific Absorption Rates (SARs), as shown in Fig. \ref{fig:tumour_results} (b), reveal distinct patterns. Without a tumor, SAR is highest near the RF source and decreases rapidly with distance. However, in the presence of a tumor, SAR initially follows a similar pattern but then exhibits a local peak at a short distance from the source before continuing to decrease. This observation indicates that tumors influence how surrounding tissue absorbs electromagnetic energy due to their unique electrical properties, which may include higher water content and distinct cellular structures. The SAR peak observed with a tumor suggests that at a specific distance from the RF source, the tumor absorbs more energy than the surrounding tissue. This difference in energy absorption characteristics may be influenced by factors such as the tumor's location, size, or composition.



\section{Conclusion}
\label{sec:conclusions} 
In this study, we explored the application of radar technology for brain imaging, employing both Patch and Vivaldi antennas. The Vivaldi antenna proved effective for deep tissue scans, while the Patch antenna demonstrated precision in targeting specific brain regions, notably in tumor detection. We also established safety guidelines based on Specific Absorption Rate (SAR) and Voltage Standing Wave Ratio (VSWR). While our findings are promising, the use of a standardized head model is a significant limitation. Future studies should incorporate a variety of head models, including those derived from actual patient scans, to enhance the generalizability of our results. Additionally, real-time data acquisition and the incorporation of dynamic brain activities in our simulations could further validate our findings. Collaborative efforts with neurologists and clinical researchers will be crucial in adapting our radar technology for clinical use. Finally, exploring the integration of machine learning algorithms for automated tumor detection could streamline diagnostic processes, making radar technology not only a viable alternative but a preferred method in certain clinical scenarios.
\bibliographystyle{ieeetr}
\bibliography{Alternate-Transactions-Brief-LaTeX-template/paper}

\begin{thebibliography}{10}

\bibitem{westerlund_stem_2003}
U.~Westerlund, M.~C. Moe, M.~Varghese, J.~Berg-Johnsen, M.~Ohlsson, I.~A. Langmoen, and M.~Svensson, ``Stem cells from the adult human brain develop into functional neurons in culture,'' {\em Experimental Cell Research}, vol.~289, pp.~378--383, Oct. 2003.

\bibitem{hossain_lightweight_2023}
A.~Hossain, M.~T. Islam, S.~K. Abdul~Rahim, M.~A. Rahman, T.~Rahman, H.~Arshad, A.~Khandakar, M.~A. Ayari, and M.~E.~H. Chowdhury, ``A {Lightweight} {Deep} {Learning} {Based} {Microwave} {Brain} {Image} {Network} {Model} for {Brain} {Tumor} {Classification} {Using} {Reconstructed} {Microwave} {Brain} ({RMB}) {Images},'' {\em Biosensors}, vol.~13, p.~238, Feb. 2023.

\bibitem{talukder_compact_2021}
M.~S. Talukder, M.~Samsuzzaman, M.~T. Islam, R.~Azim, M.~Z. Mahmud, and M.~T. Islam, ``Compact ellipse shaped patch with ground slotted broadband monopole patch antenna for head imaging applications,'' {\em Chinese Journal of Physics}, vol.~72, pp.~310--326, Aug. 2021.

\bibitem{rodriguez-duarte_brick-shaped_2020}
D.~O. Rodriguez-Duarte, J.~A.~T. Vasquez, R.~Scapaticci, L.~Crocco, and F.~Vipiana, ``Brick-{Shaped} {Antenna} {Module} for {Microwave} {Brain} {Imaging} {Systems},'' {\em IEEE Antennas and Wireless Propagation Letters}, vol.~19, pp.~2057--2061, Dec. 2020.

\bibitem{ullah_experimental_2022}
R.~Ullah, I.~Saied, and T.~Arslan, ``Experimental radar data for monitoring brain atrophy progression,'' {\em Data in Brief}, vol.~43, p.~108379, Aug. 2022.

\bibitem{hossain_brain_2023}
A.~Hossain, M.~T. Islam, T.~Rahman, M.~E.~H. Chowdhury, A.~Tahir, S.~Kiranyaz, K.~Mat, G.~K. Beng, and M.~S. Soliman, ``Brain {Tumor} {Segmentation} and {Classification} from {Sensor}-{Based} {Portable} {Microwave} {Brain} {Imaging} {System} {Using} {Lightweight} {Deep} {Learning} {Models},'' {\em Biosensors}, vol.~13, p.~302, Feb. 2023.

\bibitem{ojaroudi_pattern-reconfigurable_2021}
M.~Ojaroudi and S.~Bila, ``Pattern-{Reconfigurable} {Metasurface}-{Antenna} {Array} for {Functional} {Brain} {Imaging} {Applications},'' in {\em 2021 15th {European} {Conference} on {Antennas} and {Propagation} ({EuCAP})}, (Dusseldorf, Germany), pp.~1--5, IEEE, Mar. 2021.

\bibitem{akazzim_multi-element_2022}
Y.~Akazzim, O.~El~Mrabet, J.~Romeu, and L.~Jofre-Roca, ``Multi-{Element} {UWB} {Probe} {Optimization} for {Medical} {Microwave} {Imaging},'' {\em Sensors}, vol.~23, p.~271, Dec. 2022.

\bibitem{ghavami_use_2022}
N.~Ghavami, E.~Razzicchia, O.~Karadima, P.~Lu, W.~Guo, I.~Sotiriou, E.~Kallos, G.~Palikaras, and P.~Kosmas, ``The {Use} of {Metasurfaces} to {Enhance} {Microwave} {Imaging}: {Experimental} {Validation} for {Tomographic} and {Radar}-{Based} {Algorithms},'' {\em IEEE Open Journal of Antennas and Propagation}, vol.~3, pp.~89--100, 2022.

\bibitem{margish_s_joshi_analysis_2016}
{Margish S. Joshi}, {Gaurav R. Joshi}, and {B H Gardi College of Engineering and Technology}, ``Analysis of {SAR} induced in {Human} {Head} due to the exposure of {Non}-ionizing {Radiation},'' {\em International Journal of Engineering Research and}, vol.~V5, p.~IJERTV5IS020466, Feb. 2016.

\bibitem{gabriel_dielectric_1996-1}
S.~Gabriel, R.~W. Lau, and C.~Gabriel, ``The dielectric properties of biological tissues: {III}. {Parametric} models for the dielectric spectrum of tissues,'' {\em Physics in Medicine and Biology}, vol.~41, pp.~2271--2293, Nov. 1996.

\bibitem{hesabgar_dielectric_2017}
S.~M. Hesabgar, A.~Sadeghi-Naini, G.~Czarnota, and A.~Samani, ``Dielectric properties of the normal and malignant breast tissues in xenograft mice at low frequencies (100 {Hz}–1 {MHz}),'' {\em Measurement}, vol.~105, pp.~56--65, July 2017.

\bibitem{Chopra2015}
K.~Chopra, D.~Calva, M.~Sosin, K.~K. Tadisina, A.~Banda, C.~De~La~Cruz, M.~R. Chaudhry, T.~Legesse, C.~B. Drachenberg, P.~N. Manson, and M.~R. Christy, ``A comprehensive examination of topographic thickness of skin in the human face,'' {\em Aesthetic Surgery Journal}, vol.~35, p.~1007–1013, Oct. 2015.

\bibitem{MoreiraGonzalez2006}
A.~Moreira-Gonzalez, F.~E. Papay, and J.~E. Zins, ``Calvarial thickness and its relation to cranial bone harvest,'' {\em Plastic and Reconstructive Surgery}, vol.~117, p.~1964–1971, May 2006.

\bibitem{Fam2018}
M.~Fam, A.~Potash, M.~Potash, R.~Robinson, L.~Karnell, E.~O’Brien, and J.~Greenlee, ``Skull base dural thickness and relationship to demographic features: A postmortem study and literature review,'' {\em Journal of Neurological Surgery Part B: Skull Base}, vol.~79, p.~614–620, June 2018.

\bibitem{Haeussinger2011}
F.~B. Haeussinger, S.~Heinzel, T.~Hahn, M.~Schecklmann, A.-C. Ehlis, and A.~J. Fallgatter, ``Simulation of near-infrared light absorption considering individual head and prefrontal cortex anatomy: Implications for optical neuroimaging,'' {\em PLoS ONE}, vol.~6, p.~e26377, Oct. 2011.

\bibitem{Winkler2010}
A.~M. Winkler, P.~Kochunov, J.~Blangero, L.~Almasy, K.~Zilles, P.~T. Fox, R.~Duggirala, and D.~C. Glahn, ``Cortical thickness or grey matter volume? the importance of selecting the phenotype for imaging genetics studies,'' {\em NeuroImage}, vol.~53, p.~1135–1146, Nov. 2010.

\bibitem{Schnack2014}
H.~G. Schnack, N.~E.~M. van Haren, R.~M. Brouwer, A.~Evans, S.~Durston, D.~I. Boomsma, R.~S. Kahn, and H.~E. Hulshoff~Pol, ``Changes in thickness and surface area of the human cortex and their relationship with intelligence,'' {\em Cerebral Cortex}, vol.~25, p.~1608–1617, Jan. 2014.

\bibitem{zhang_microwave_2012}
H.~Zhang, B.~Flynn, A.~T. Erdogan, and T.~Arslan, ``Microwave imaging for brain tumour detection using an {UWB} {Vivaldi} {Antenna} array,'' in {\em 2012 {Loughborough} {Antennas} \& {Propagation} {Conference} ({LAPC})}, (Loughborough, Leicestershire, United Kingdom), pp.~1--4, IEEE, Nov. 2012.

\end{thebibliography}
\end{document}